\documentclass[journal,twoside,web]{ieeecolor}
\usepackage{generic}
\usepackage{cite}
\usepackage{amsmath,amssymb,amsfonts}
\usepackage{graphicx}
\usepackage{textcomp}
\usepackage{multirow}
\usepackage{algorithm}
\usepackage{algpseudocode}

\algrenewcommand\algorithmicrequire{\textbf{Input:}}
\algrenewcommand\algorithmicensure{\textbf{Output:}}

\def\BibTeX{{\rm B\kern-.05em{\sc i\kern-.025em b}\kern-.08em
    T\kern-.1667em\lower.7ex\hbox{E}\kern-.125emX}}
\begin{document}
\title{A Unified Multi-Phase CT Synthesis and Classification Framework for Kidney Cancer Diagnosis with Incomplete Data  
}
\author{Kwang-Hyun Uhm, Seung-Won Jung, \IEEEmembership{Senior Member, IEEE}, Moon Hyung Choi, Sung-Hoo Hong, and Sung-Jea Ko, \IEEEmembership{Fellow, IEEE}
\thanks{This work was supported by the Korea Medical Device Development Fund grant funded by the Korea government (the Ministry of Science and ICT, the Ministry of Trade, Industry and Energy, the Ministry of Health \& Welfare, the Ministry of Food and Drug Safety) (Project Number: RS-2020-KD000096. \textit{(Corresponding author: Seung-Won Jung.)}}
\thanks{
Kwang-Hyun Uhm, Seung-Won Jung, and Sung-Jea Ko are with the Department of Electrical Engineering, Korea University, Seongbuk-gu, Seoul 02841, South Korea (e-mail: khuhm@dali.korea.ac.kr; swjung83@korea.ac.kr; sjko@korea.ac.kr).}
\thanks{
Moon-Hyung Choi is with the Department of Radiology, The Catholic University of Korea, Seoul, South Korea (e-mail: choimh1205@gmail.com).}
\thanks{
Sung-Hoo Hong is with the Department of Urology, The Catholic University of Korea, Seoul, South Korea (e-mail: toomey@catholic.ac.kr).}}

\maketitle

\begin{abstract}
Multi-phase CT is widely adopted for the diagnosis of kidney cancer due to the complementary information among phases. However, the complete set of multi-phase CT is often not available in practical clinical applications. In recent years, there have been some studies to generate the missing modality image from the available data. Nevertheless, the generated images are not guaranteed to be effective for the diagnosis task. In this paper, we propose a unified framework for kidney cancer diagnosis with incomplete multi-phase CT, which simultaneously recovers missing CT images and classifies cancer subtypes using the completed set of images. The advantage of our framework is that it encourages a synthesis model to explicitly learn to generate missing CT phases that are helpful for classifying cancer subtypes. We further incorporate lesion segmentation network into our framework to exploit lesion-level features for effective cancer classification in the whole CT volumes. The proposed framework is based on fully 3D convolutional neural networks to jointly optimize both synthesis and classification of 3D CT volumes. Extensive experiments on both in-house and external datasets demonstrate the effectiveness of our framework for the diagnosis with incomplete data compared with state-of-the-art baselines. In particular, cancer subtype classification using the completed CT data by our method achieves higher performance than the classification using the given incomplete data.

\end{abstract}


\section{Introduction}
\label{Introduction}
Cancer subtype classification is a crucial step in patient management, as treatment planning and prognosis prediction are dependent on pathological subtype of tumors~\cite{kim2002rcc}. For kidney cancer, there are five major subtypes of renal tumors: clear cell renal cell carcinoma (ccRCC), papillary renal cell carcinoma (pRCC), chromophobe renal cell carcinoma (chRCC), angiomyolipoma (AML), and oncocytoma. 
Medical imaging is widely used for the non-invasive diagnosis of cancer, which can prevent unnecessary biopsy or surgery~\cite{wang2021rcc, han2019rcc}. Typically, multi-modal medical images are required to accurately diagnose patients since they provide complementary visual information about lesions. For example, four-phase dynamic contrast-enhanced computed tomography (CT), which captures a series of CT volumes before and after contrast injection at different time points, is used for differential diagnosis of kidney cancer~\cite{young2013ccrcc}.
Multi-parametric magnetic resonance imaging (MRI) is used for brain disease diagnosis~\cite{drevelegas2010brain}. 

However, the complete set with all modalities is often not available in clinical practice due to different imaging protocols among medical institutions, acquisition cost, image corruption, and patient characteristics~\cite{CHASE}. In addition, motion artifacts caused by breathing, allergic reactions to the contrast material, and systematic error in scanners also lead to phase missing~\cite{breath}. Specifically, kidney cancer is usually asymptomatic and found incidentally during screening for unrelated indications~\cite{incidental}, which may require less than four CT phases. It is very common to perform two-phase or three-phase CT examination for routine assessment in clinical practice. As repetition of CT examination to acquire the four-phase CT for kidney cancer diagnosis is undesirable due to the additional cost and radiation exposure~\cite{radiation, cost}, radiologists often diagnose lesions with the available CT~\cite{young2013ccrcc}. However, as differential diagnosis of kidney cancer is challenging due to subtle differences in image features of renal tumors, more CT phases can benefit the  diagnosis. 
Therefore, it would be useful to generate missing CT phase without repetition of CT imaging for more accurate preoperative diagnosis.

The incomplete data cannot be directly applied to multi-modal analysis algorithms, which require the complete dataset as an input. Some valuable information in the collected dataset cannot be used if a diagnosis model can be trained using only the available complete data. To deal with this issue, one can extract features from each modality and then aggregate features from multiple modalities to predict class labels~\cite{su2015multi}. Yuan \textit{et al.}~\cite{YUAN2012} proposes a multi-source feature learning for brain disease classification with incomplete multiple heterogeneous neuroimaging data.
Zhou \textit{et al.}~\cite{Zhou2019} proposes a latent representation learning method utilizing the inter-modality associations for alzheimer’s disease diagnosis with incomplete multi-modality data~\cite{Zhou2019}.
On the other hand, imputation of missing modality can be an effective strategy to better utilize all available data for developing an effective multi-modal diagnosis system~\cite{Lee2019collagan}.

Recently, generative adversarial networks (GANs) have shown promising results on various image synthesis tasks including medical image synthesis, aiming at producing images that are indistinguishable from real images~\cite{choi2018stargan, goodfellow2014gan, isola2017pix2pix, Lee2019collagan, seo2021neural, zhu2017cyclegan}. There have been many studies in the literature to translate images from one domain to another~\cite{choi2018stargan, isola2017pix2pix, zhu2017cyclegan}. These methods, however, do not take into account the input images from multiple domains. As a more advanced model, CollaGAN~\cite{Lee2019collagan} is designed for missing MRI contrast synthesis in which the missing image is generated from all available images using the multiple cycle consistency loss, achieving a higher visual quality than other approaches. Zhang \textit{et al.} \cite{zhang2018trans} employ a shape consistency loss, which is supervised by segmentation networks, to maintain anatomical structures in synthesized medical images, leading to improved segmentation performance. Although existing methods successfully learn to generate plausible images, there is no guarantee that the generated images contain diagnostically meaningful visual information, especially for cancer subtype classification. This limits the practical applicability of the existing synthesis models to the clinical diagnosis with incomplete data. 

\begin{figure}[!t]
\centering
\includegraphics[width=1.0\linewidth]{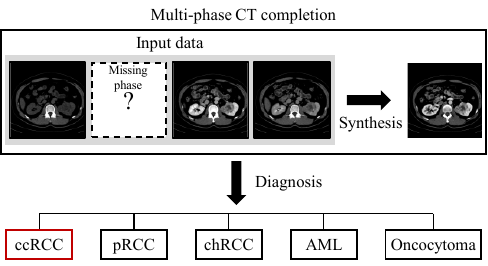}
\caption{Illustration of the proposed joint CT synthesis and cancer subtype classification pipeline. The missing phase is synthesized to accurately diagnose cancer using the completed multi-phase CT. The diagnosed subtype label is highlighted in red.}
\label{fig:illustration}
\end{figure}

In this paper, we propose a GAN-based framework for diagnostically informative image synthesis, called DiagnosisGAN, which jointly learns missing CT phase synthesis and cancer subtype classification in a unified framework, as illustrated in Fig.~\ref{fig:illustration}. Our key idea is to learn how to generate missing CT images such that they are effective for classifying pathological subtypes of tumor, while learning cancer subtype classification from completed multi-phase CT with synthesized images. 
Specifically, the training of the missing phase generator is explicitly supervised by cancer subtype classification to encourage the generator to be aware of meaningful features for differentiating cancer subtypes in CT images while learning to synthesize images.
DiagnosisGAN is designed based on 3D convolutional neural networks (CNNs) to synthesize and classify 3D CT volumes. In addition, since the classification of cancer subtypes directly from the whole CT volume is difficult due to the small tumors, we extract the lesion-level features using a pretrained 3D lesion segmentation network for the subtype classification. In our framework, CT phase synthesis and cancer subtype classification are jointly optimized to further improve the diagnostic performance. We observe that the performance of cancer subtype classification with the completed multi-phase CT by the proposed model can achieve higher performance than the classification with the given incomplete CT in our experiments.

Our main contributions are summarized as follows.
\begin{itemize}
  \item To the best of our knowledge, this is the first work to generate missing CT phases to be helpful for diagnosis, especially for cancer subtype classification.
  \item We propose a unified framework for multi-phase CT synthesis and cancer subtype classification, namely DiagnosisGAN, where both tasks are jointly optimized.
  \item We design DiagnosisGAN using fully 3D CNNs for effective volumetric CT image synthesis, lesion segmentation, and classification. 
  \item Extensive experiments on in-house and external multi-phase CT scans of kidney cancer patients demonstrate the effectiveness of DiagnosisGAN on both CT synthesis and subtype classification, compared to the baselines derived from state-of-the-art approaches.
\end{itemize}

\begin{figure*}
\begin{center}
\includegraphics[width=1.0\linewidth]{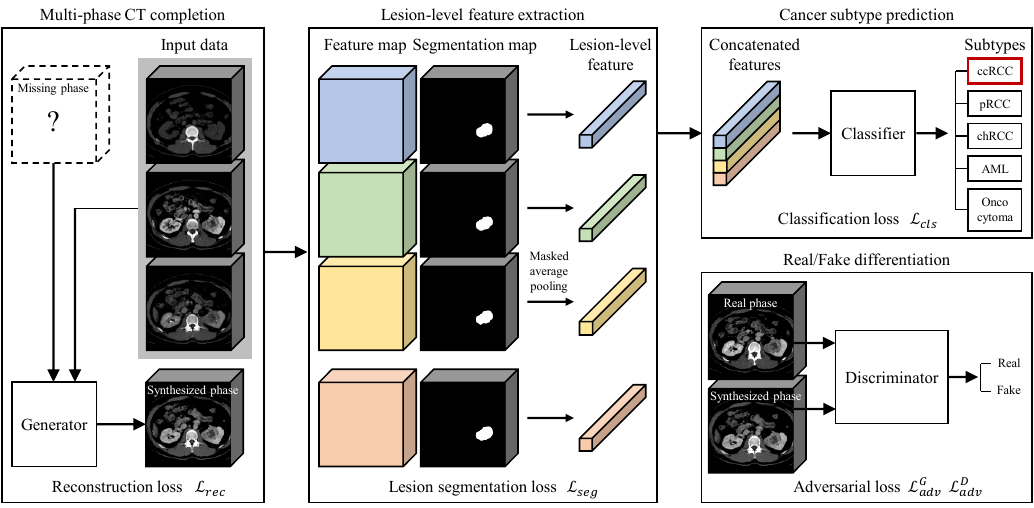}
\end{center}
   \caption{Overall framework of our DiagnosisGAN. 
   The generator takes an incomplete multi-phase CT data as input and produces the synthesized volume for the missing phase, whereas the discriminator tries to differentiate the real and synthesized phases. For each volume in the completed CT, the feature map and the tumor segmentation map are extracted, and the lesion-level feature is obtained via masked average pooling. Finally, the classifier takes the concatenated lesion-level features as input and outputs the cancer subtype prediction. 
   }
\label{fig:framework}
\end{figure*}

\section{Related work}
\label{Related work}
\noindent \textbf{Generative Adversarial Network.}
GANs~\cite{goodfellow2014gan} have been widely used in many image-to-image translation tasks and shown great success in generating images that are indistinguishable from real images~\cite{choi2018stargan, choi2020starganv2, huang2018munit, isola2017pix2pix, kim2017discogan, Lee2019collagan, zhu2017cyclegan, zhu2017bicyclegan}. Pix2Pix~\cite{isola2017pix2pix} presented a general solution of image-to-image translation using conditional GANs. CycleGAN~\cite{zhu2017cyclegan} introduced a cycle consistency loss to learn the mapping between two different domains. StarGAN~\cite{choi2018stargan} and RadialGAN~\cite{yoon2018radialgan} performed image translation across multiple domains using a single generator with domain information. These methods focused on translating images from one domain to another without considering the images in the other existing domains. Recently, CollaGAN~\cite{Lee2019collagan} was proposed to utilize a multiple set of input images to impute the missing domain image by using the multiple cycle consistency loss, while retaining the single generator.

\noindent \textbf{Medical Image Synthesis.}
With the advancement of GANs, medical image synthesis has been extensively studied for the data augmentation~\cite{FRIDADAR2018321, focal_aug, lung_aug}, generation of missing MRI~\cite{dar2019mri, Lee2019collagan, sharma2020mri, yu2019eagan, tmi2021comp} or CT~\cite{liu2020dye, seo2021neural}, and for translation between MRI and CT~\cite{huo2019synseg, nie2017mri_ct, Wolterink2017mri_ct, zhang2018trans}. In~\cite{FRIDADAR2018321, focal_aug, lung_aug}, synthetic CT images are generated to augment training data for classification task, but these methods do not address the problem of missing data reconstruction from incomplete multi-modal data.
Seo \textit{et al.} \cite{seo2021neural} and Liu \textit{et al.} \cite{liu2020dye} applied a GAN for synthesizing contrast-enhanced CT images from given non-contrast CT images. CollaGAN~\cite{Lee2019collagan} generated one missing MRI image using the other three contrast images. Zhang \textit{et al.} \cite{zhang2018trans} and Huo \textit{et al.} \cite{huo2019synseg} proposed an end-to-end synthesis and segmentation network to ensure consistent anatomical structures in synthesized volumes. ReMIC~\cite{tmi2021comp} utilizes representational disentanglement strategy which decomposes images into a shared content structure and separate styles across modalities. Although these methods achieved impressive results in synthesizing visually plausible images, they are not guaranteed to produce diagnostically informative images, especially for differential diagnosis of tumors.

\noindent \textbf{Cancer Subtype Classification.}
Classifying cancer subtypes is an important task since treatment planning and prognosis prediction depend on the pathological subtype of tumor~\cite{young2013ccrcc}. Multi-phase dynamic contrast-enhanced CT is often used for non-invasive diagnosis of cancer due to complementary visual features across phases~\cite{Pierorazio2013multi_ct, Schieda2020multi_ct, You2019multi_ct, young2013ccrcc}. Recently, CNN-based cancer subtype classification on multi-phase CT has been actively explored for liver lesions~\cite{cao2020liver, huo2020liver, liang2018liver, zhou2021liver} and renal masses~\cite{coy2019renal, han2019rcc, oberai2020renal, tanaka2020renal, deepkidney}.
Huo \textit{et al.} \cite{huo2020liver} proposed an automated detection and classification framework to differentiate four major liver lesion types using four-phase CT. Oberai \textit{et al.} \cite{oberai2020renal} investigated the ability of CNNs to classify renal masses into benign and malignant while also using four-phase CT. In the work of ~\cite{deepkidney}, the performance of CNNs in classifying five major subtypes of renal tumors is investigated and compared with six radiologists to explore the clinical applicability. However, previous approaches operate only on the complete four-phase CT data~\cite{cao2020liver, huo2020liver, oberai2020renal} or the pre-defined combination of three phases~\cite{han2019rcc, liang2018liver, zhou2021liver}, hindering their use in the clinical environment. Moreover, most of the existing methods are based on 2D CNNs which cannot fully take advantage of 3D information of CT volumes. We aim at improving the diagnostic performance when incomplete data is given, by diagnostically helpful image synthesis.

\section{Method}
\label{sec:Method}
For brevity, we explain our method for the case where there is one missing phase in the multi-phase CT scan, but our method is not restricted to this case and can be extended to the cases with two or more missing phases. Our goal is then to generate the missing phase CT image $\hat{I}_m$ from the given incomplete image set $\mathcal{I}_m=\{I_i\}_{i=1, i\neq m}^N$ , where $1\leq m\leq N$ and $N$ is the number of possible phases, such that we can accurately predict the patient’s cancer subtype $\hat{p}$ using the completed image set $\mathcal{I}_m^+=\mathcal{I}_m\cup \{\hat{I}_m\}$, where $\hat{p}$ denotes the predicted probability distribution over subtype classes. To achieve this, we jointly learn a missing phase generator $G: \mathcal{I}_m\mapsto \hat{I}_m$ and a cancer subtype classifier $C: \mathcal{I}_m^+\mapsto \hat{p}$ in a unified framework. That is, DiagnosisGAN optimizes the generator $G$ to minimize the error of the classifier $C$. The overall framework of the proposed method is illustrated in Fig.~\ref{fig:framework}. The details of our framework and loss functions are described in the following.

\begin{figure}
\begin{center}
\includegraphics[width=1.0\linewidth]{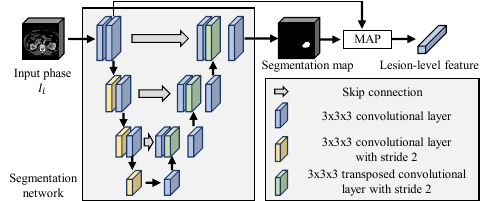}
\end{center}
   \caption{
   The architecture of the segmentation network $S$. The feature map from the first two convolutional layers is used to extract the lesion-level feature. MAP represents the masked average pooling.
   }
\label{fig:architecture}
\end{figure}

\subsection{Framework}
\noindent \textbf{Multi-phase CT synthesis.}
The generator $G$ synthesizes the missing image ${\hat{I}}_m$ by taking the incomplete image set $\mathcal{I}_m$ as input. We adopt the 3D U-Net architecture~\cite{cicek2016unet3d} which consists of $3\times3\times3$ convolutional layers, strided convolutions, trilinear up-sampling layers, and skip connections, as the generator $G$ to synthesize the whole 3D CT volume. We use the trilinear interpolation for up-sampling instead of transposed convolutions since they tend to produce checkboard artifacts~\cite{odena2016check} in synthesized volumes. The target phase label $m$ is fed to the generator $G$ with the input data $\mathcal{I}_m$ in the form of an $N$-channel binary mask $\mathbf{m}$ whose values are all ones for the missing phase and all zeros otherwise, which was introduced in CollaGAN~\cite{Lee2019collagan}. To be more specific, we concatenate images along channel dimension while the entry of missing phase is filled with zeros, which can be expressed as $\tilde{\mathcal{I}}=\left[I_1,I_2, \mathbf{0}, \ldots,I_N\right]$, where $\mathbf{0}$ is a zero tensor with the shape of $I_{i\neq m}$ and $\left[\cdot,\cdot\right]$ denotes concatenation along channel dimension. The concatenated images $\tilde{\mathcal{I}}$ are then combined with the mask $\mathbf{m}$, represented as $\left[\tilde{\mathcal{I}}, \mathbf{m}\right]$, and fed to the generator $G$. By explicitly represents the index of missing phase in the form of mask vector can help the network more easily understand which phase is missing. We use a discriminator $D$ to distinguish the generated image ${\hat{I}}_m=G\left(\mathcal{I}_m\right)$ from the real image $I_m$. An extended 3D version of PatchGAN~\cite{isola2017pix2pix}, which uses strided 3D convolutions, is utilized for the discriminator $D$ to determine whether a local 3D patch is real or synthesized, where $D(I)$ represents the probability of input image $I$ being a real image.

\noindent \textbf{Cancer subtype classification.}
Given the completed multi-phase CT image set $\mathcal{I}_m^+$, the classifier $C$ predicts the cancer subtype $\hat{p}$. For accurate cancer subtype classification, lesion features from four CT phases should be jointly considered. Since classifying cancer subtypes directly from the whole CT volumes is challenging due to the small tumor size, we extract lesion-level features using a 3D tumor segmentation network $S$, which is also based on 3D U-Net but with transposed convolutions for up-sampling, as shown in Fig. 3. Given an image $I\in\mathcal{I}_m^+$, the segmentation network $S$ outputs the segmentation map $\hat{s}=S\left(I\right)$, where the segmentation map $\hat{s}$ represents the predicted probability of each voxel belonging to a tumor area, and the low-level feature map $F$ is extracted by the first two convolutional layers of the segmentation network $S$. We then compute the lesion-level feature $f$ by applying the masked average pooling~\cite{PANet, SGOne} to the feature map $F$ with the segmentation map $\hat{s}$ as follows:
\begin{equation}\label{eqa:MAP}
f^k=\frac{\sum_{x\in\mathcal{X}}{F_x^k{\hat{s}}_x}}{\sum_{x\in\mathcal{X}}{\hat{s}}_x},
\end{equation}
where $\mathcal{X}$ is the set of all 3D spatial locations and $k\in\left\{1,\ 2,\ \ldots,K\right\} $, where $K$ is the dimension of the feature map $F$. We adopt the low-level features instead of higher-level ones since local textures are more helpful than global volume statistics for characterizing lesions. After obtaining the lesion-level features for all $I\in\mathcal{I}_m^+$, we concatenate them together to form an $NK$-dimensional feature vector $\mathcal{F}=\left[f_1,f_2,\ldots,f_N\right]$, where $f_i$ is the obtained lesion-level feature for the $i^{th}$ image in $\mathcal{I}_m^+$. We use it to produce the final cancer subtype prediction $\hat{p}=C(\mathcal{F})$ through several fully connected layers in the classifier $C$. We simply rewrite this formulation as $\hat{p}=C(\mathcal{I}_m^+)$, since the segmentation network $S$ is fixed during the training of DiagnosisGAN. Our approach enables the cancer subtype classifier $C$ to focus on lesion-related features of the image set $\mathcal{I}_m^+$.

\begin{algorithm}[!t]
\caption{Training algorithm of DiagnosisGAN}\label{alg:train_proc}
\begin{algorithmic}[1]
\Require Training image sets with labels $\{\mathfrak{I}, \mathfrak{s}, \mathfrak{p}\}$ and the number of iterations of initial training $T_1$ and joint fine-tuning $T_2$.
\Ensure Generator $G$, discriminator $D$, segmentor $S$, and classifier $C$ parameterized by $\theta_G$, $\theta_D$, $\theta_S$, and $\theta_C$, respectively.

\State Initialize $\theta_G$, $\theta_D$, $\theta_S$, and $\theta_C$ randomly.

\State Train $S$ by minimizing $\mathcal{L}_{seg}$ using CT images in $\mathfrak{I}$ and segmentation masks in $\mathfrak{s}$. ($\theta_S$ is updated and fixed)

\State Pre-train $C$ using CT image sets $\mathcal{I}\in\mathfrak{I}$, trained $S$, and cancer labels $p\in\mathfrak{p}$:
\[
{\theta_C} \gets {\arg \min}_{\theta_C} \mathbb{E}_{\mathcal{I},p}\left[-\log{C\left(\mathcal{I};p\right)}\right]
\]

\State Generate incomplete image sets $\mathcal{I}_m$ from the complete image sets $\mathcal{I}\in\mathfrak{I}$ by randomly dropping phases.

\For{$t \gets 1$ to $T_1$}

Compute discriminator loss $\mathcal{L}_{adv}^D$ and update $\theta_D$

Compute full objective $\mathcal{L}$ and update $\theta_G$

\EndFor


\For{$t \gets 1$ to $T_2$}

Compute discriminator loss $\mathcal{L}_{adv}^D$ and update $\theta_D$

Compute classification loss $\mathcal{L}_{cls}$ and update $\theta_C$

Compute full objective $\mathcal{L}$ and update $\theta_G$

\EndFor
\end{algorithmic}
\end{algorithm}

\subsection{Loss functions}
\noindent \textbf{Adversarial loss.}
Given an incomplete image set $\mathcal{I}_m$, while $\mathcal{I}\in\mathfrak{I}$ is the complete image set from the training image dataset $\mathfrak{I}$, 
the generator $G$ learns to synthesize the missing phase image ${\hat{I}}_m=G\left(\mathcal{I}_m\right)$ that is indistinguishable by the discriminator $D$, while the discriminator $D$ learns to differentiate between the generated image $G\left(\mathcal{I}_m\right)$ and the real image $I_m$.
This min-max problem is solved by using the least-square adversarial loss~\cite{LSGAN}, which showed better training stability than the original GAN loss, defined as
\begin{equation}\label{eqa:L_adv_G}
\mathcal{L}_{adv}^G=\mathbb{E}_{\mathcal{I}_m}[\left(D\left(G\left(\mathcal{I}_m\right)\right)-1\right)^2],
\end{equation}
\begin{equation}\label{eqa:L_adv_D}
\mathcal{L}_{adv}^D=\mathbb{E}_{I_m}[\left(D\left(I_m\right)-1\right)^2]+\mathbb{E}_{\mathcal{I}_m}[\left(D\left(G\left(\mathcal{I}_m\right)\right)\right)^2].
\end{equation}
The generator $G$ and the discriminator $D$ aim to minimize $\mathcal{L}_{adv}^G$ and $\mathcal{L}_{adv}^D$, respectively.

\noindent \textbf{Reconstruction loss.}
A classical reconstruction loss is used to guide the generator $G$, which is the voxel-wise $L1$ distance between the generated image $G\left(\mathcal{I}_m\right)$ and the target image $I_m$, represented as 
\begin{equation}\label{eqa:L_rec}
\mathcal{L}_{rec}=\mathbb{E}_{\mathcal{I}_m,I_m}[\|I_m-G(\mathcal{I}_m)\|_1].
\end{equation}
Training the generator $G$ with only the reconstruction loss $\mathcal{L}_{rec}$ leads to blurry results.

\noindent \textbf{Segmentation loss.}
We use a segmentation loss to constrain the generated image $G\left(\mathcal{I}_m\right)$ to have a well-defined tumor structure. Since structures of tumor in CT images contain diagnostic information, preserving them while synthesizing missing phase is helpful for the subsequent tumor analysis task. Dice loss~\cite{Dice} is adopted instead of a cross-entropy loss used in~\cite{huo2019synseg, zhang2018trans}, which quantifies the volume overlap between the predicted segmentation map $\hat{s}=S\left(G\left(\mathcal{I}_m\right)\right)$ and the ground-truth segmentation map $s\in\mathfrak{s}$, where $\mathfrak{s}$ is a set of annotated segmentation masks for training images, defined as

\begin{equation}\label{eqa:L_seg}
\mathcal{L}_{seg}=\mathbb{E}_{\hat{s}, s}\left[1-\frac{2\sum_{x\in\mathcal{X}}{s_x{\hat{s}}_x}}{\sum_{x\in\mathcal{X}}s_x^2+\sum_{x\in\mathcal{X}}{\hat{s}}_x^2}\right].
\end{equation}

\noindent \textbf{Cancer subtype classification loss.}
Given the completed image set $\mathcal{I}_m^+$, the classifier $C$ learns to predict the cancer subtype $\hat{p}=C\left(\mathcal{I}_m^+\right)$. 
We use the cross-entropy loss for the training of the classifier $C$, which is defined as
\begin{equation}\label{eqa:L_cls}
\mathcal{L}_{cls}=\mathbb{E}_{\mathcal{I}_m^+,p}\left[-\log{C\left(\mathcal{I}_m^+;p\right)}\right],
\end{equation}
where $p\in\mathfrak{p}$ is the true subtype class, $\mathfrak{p}$ is a set of cancer subtype labels for training cases, and $C\left(\mathcal{I}_m^+;p\right)$ represents the predicted probability for the target class $p$.

\noindent \textbf{Full objective.}
The goal of DiagnosisGAN is to generate diagnostically informative images that are helpful for cancer subtype classification. 
Toward this goal, we jointly learn the generator $G$ and the classifier $C$ such that the generator $G$ tries to minimize the error of the cancer subtype classifier $C$.
Thus, our full objective can be formulated as

\begin{equation}\label{eqa:L_full}
\mathcal{L}=\mathcal{L}_{adv}^G+\lambda_{rec}\mathcal{L}_{rec}+\lambda_{seg}\mathcal{L}_{seg}+\lambda_{cls}\mathcal{L}_{cls},
\end{equation}
where $\lambda_{rec}$, $\lambda_{seg}$, and $\lambda_{cls}$ are the weights to balance different loss terms.
The generator $G$ is trained to minimize this full objective $\mathcal{L}$, while the classifier $C$ tries to minimize $\mathcal{L}_{cls}$.
This joint optimization encourages the generator $G$ to produce the missing phase images that can lead to accurate predictions of cancer subtypes by the classifier $C$. The overall training procedure is summarized in Alg.~\ref{alg:train_proc}, and more details are described in Section~\ref{sec:Implementation}.


\begin{table}[!t]
\caption{Patient demographics, phase number, subtype and tumor size distributions for training/testing dataset.}
\centering
\begin{tabular}{|l|c|c|c|c|}
\hline 
\multirow{3}{8em}{Patients ($n$)} & \multicolumn{3}{|c|}{In-house dataset} & \multirow{2}{*}{TCIA}\\
\cline{2-4}
 & Total & Training & Testing & \\
\cline{2-5}
 & 249 & 146 & 103 & 139\\
\hline\hline
\textbf{Age (years)} & & & &\\
--40 & 33 & 19 & 14 & 13\\
40--50 & 58 & 39 & 19 & 26\\
50--60 & 70 & 36 & 34 & 41\\
60--70 & 57 & 33 & 24 & 30\\
70-- & 31 & 19 & 12 & 29\\
\hline
\textbf{Gender} & & & &\\
Female & 113 & 63 & 50 & 37\\
Male & 136 & 83 & 53 & 102\\
\hline
\textbf{Phase number} & & & &\\
Four-phase & 183 & 146 & 37 & -\\
Three-phase & 66 & - & 66 & 139\\
\hline
\textbf{Subtype} & & & &\\
ccRCC & 64 & 50 & 14 & 119\\
pRCC & 53 & 26 & 27 & 14\\
chRCC & 55 & 29 & 26 & 6\\
AML & 48 & 25 & 23 & -\\
Oncocytoma & 29 & 16 & 13 & -\\
\hline
\textbf{Tumor Size (cm)} & & & &\\
1--2 & 64 & 34 & 30 & \multirow{6}{*}{-}\\
2--3 & 67 & 46 & 21 & \\
3--4 & 40 & 27 & 13 & \\
4--5 & 29 & 17 & 12 & \\
5--6 & 27 & 14 & 13 & \\
6--7 & 22 & 8 & 14 & \\
\hline
\end{tabular}
\label{tab:demographics}
\end{table}

\section{Experiments and Discussion}
\subsection{Dataset}\label{sec:dataset}
\noindent \textbf{In-house dataset.} 
We collect 249 multi-phase dynamic contrast-enhanced CT scans of kidney cancer patients who underwent nephrectomy for renal tumors in Seoul St. Mary's Hospital. This study was approved by the Seoul St. Mary’s Hospital Institutional Review Board. The patients who underwent abdominal CT scans within three months before surgery are included. Among the 249 patients, 183 patients underwent complete four-phase CT scans (732 volumes), while 66 patients underwent CT scans with three phases (198 volumes). Our dataset contains five major subtypes of renal tumors: ccRCC, pRCC, chRCC, AML, and oncocytoma. The subtype labels of all tumors are pathologically confirmed by surgical resection. Although AML and oncocytoma are benign tumors, they are often mistaken for malignant ones due to the substantial overlap in imaging features, which may lead to unnecessary biopsy or surgery~\cite{EAU}. The diameter of the tumor ranges from 1 to 7 cm. There are four different contrast phases in each patient's CT data: non-contrast, arterial, portal, and delayed phases. These phases are obtained at different time points after injecting the contrast material. From the 183 complete CT cases, 37 patient cases are randomly selected for the testing and the rest are used for the training. From the training set, 29 cases are randomly selected and used for validation. As a result, the dataset is divided into train/val/test split of 65\%/15\%/20\%.  For each complete four-phase CT scan, we generate four simulated incomplete CT scans by dropping one of the phase, each of which has a different missing phase (non-contrast, arterial, portal, and delayed missing phase), to learn to synthesize images of various phases. In addition, 66 incomplete CT cases are used as real-world test cases. Dataset characteristics of training and testing sets are summarized in Table~\ref{tab:demographics}. Voxel-level segmentation masks for kidney and tumor in each CT volume are annotated for training the segmentation network $S$. Since there is a large number of slices in CT volumes to be annotated, 10 trained annotators first delineate the kidneys and tumors, and these annotations are checked and refined by an experienced radiologist (11 years).

\noindent \textbf{TCIA.} 
For external validation, we collect 139 incomplete three-phase CT scans from The Cancer Imaging Archive (TCIA)~\cite{TCIA}, which is a large public database of cancer images. This dataset contains three RCC subtype classes (ccRCC, pRCC, and chRCC). The benign cases (AML and oncocytoma) are not available in TCIA. The characteristics of the dataset are presented in Table~\ref{tab:demographics}. We evaluate the performance on this external test dataset for the models trained on our in-house dataset to demonstrate the generalizability of our method.   

\noindent \textbf{Data preprocessing.} 
The size of each slice in a CT volume is 512$\times$512, and the number of slices varies across scans. The pixel spacing ranges from 0.53 to 0.94 mm, and the slice thickness ranges from 1 to 7 mm. To deal with varying voxel spacings, we resample all volumes to 1.5$\times$1.5$\times$3 mm$^3$ voxel size. We clip the intensity values of voxels to [-40, 350] HU, which covers the intensity range of kidneys and tumors, and normalize the values by the mean and standard deviation calculated from the training dataset. Since there are some spatial misalignments between phases due to the patient movements, all phases are registered using DEEDS~\cite{DEEDS}.

\subsection{Implementation details}\label{sec:Implementation}
We implement DiagnosisGAN in PyTorch~\cite{Pytorch} and train it with an NVIDIA TITAN Xp GPU. We crop 192$\times$160$\times$96 regions from the whole volume containing kidneys and tumors for computation efficiency. We utilize the 3D U-Net architecture~\cite{cicek2016unet3d} for the generator $G$ and the segmentation network $S$, where each 3$\times$3$\times$3 convolutional layer is followed by instance normalization (IN)~\cite{IN} and LeakyReLU~\cite{LeakyRelu} activation. Downsampling is performed using convolutional layers with stride 2, and transposed convolution with stride 2 is used for upsampling in the segmentation network $S$, while trilinear interpolation is used in the generator $G$. The discriminator $D$ is composed of a series of $4\times4\times4$ convolutional layers with stride of 2, each followed by IN and LeakyReLU. Motivated by PatchGAN~\cite{isola2017pix2pix}, we use a local 88$\times$88$\times$43 volume patch for discriminating whether it is real or fake. The classifier $C$ consists of three fully connected layers with ReLU activation, and a softmax layer to produce the probability distribution. We apply dropout on the first two fully connected layers with a ratio of 0.5 to prevent overfitting. The parameters in (\ref{eqa:L_full}) are tuned by cross-validation, and obtained as follows: $\lambda_{rec}=1$, $\lambda_{seg}=0.1$, and $\lambda_{cls}=0.1$.


The whole training procedure of DiagnosisGAN takes the following steps. First, we train the segmentation network $S$ using Dice loss~\cite{Dice} with stochastic gradient descent. We set the batch size to 2 and the initial learning rate to 0.01, and the learning rate is decreased by a polynomial decay with a power of 0.9. Data augmentations including image rotation, scaling, elastic deformation, flipping, and Gaussian noise addition are applied. Then, we train the classifier $C$ on complete CT data with the help of the trained segmentation network $S$ using the classification loss $\mathcal{L}_{cls}$, which is minimized by Adam optimizer~\cite{Adam} with the learning rate of 0.0001 and the batch size of 1. Next, we train the generator $G$ using the full objective $\mathcal{L}$ with the fixed classifier $C$ and train the discriminator $D$ using $\mathcal{L}_{adv}^D$. Adam optimizer with the learning rate of 0.0001 is employed, and the batch size is set to 1. Finally, we jointly fine-tune the generator $G$ with the full objective $\mathcal{L}$ and the classifier $C$ with $\mathcal{L}_{cls}$ using the dataset completed by $G$. 


\begin{figure*}[!t]
\centering
\includegraphics[width=1.0\linewidth]{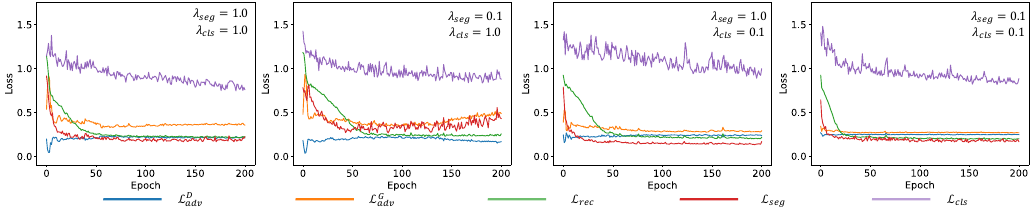}
\caption{The loss curves of DiagnosisGAN for training on different hyper-parameter settings.}
\label{fig:loss_curves_abl}
\end{figure*}

\begin{table*}
\caption{Quantitative comparison of cancer subtype classification performances with 37$\times$4 (simulated) incomplete multi-phase CT data.}
\begin{center}
\begin{tabular}{|l|l|c c c c c|c|c|}
\hline 
Classifier input & Method & ccRCC & pRCC & chRCC & AML & Oncocytoma & mAUC & $p$-value\\
\hline\hline
Three-phase & Cls-3P & \textbf{74.8} & 85.5 & 88.2 & \textbf{89.5} & 69.3 & 81.5 & 0.0473\\
\hline
\multirow{6}{7em}{Three-phase + One synthesized} & BaseSyn  & 73.5 & 84.4 & 88.5 & 86.7 & 72.3 & 81.1 & 0.0177   \\
 & CollaGAN  & 71.0 & 84.5 & 88.6 & 86.6 & 73.7 & 80.9 & 0.0154 \\
 & Syn-Seg & 70.9 & 84.2 & 90.3 & 87.6 & 73.5 & 81.3 & 0.0405\\
 &ReMIC & 72.9 & 84.2 & 89.5 & 87.7 & 70.9 & 81.0 &  0.0273 \\
\cline{2-9}
& DiagnosisGAN (w/o joint) & 73.6 & \textbf{86.8} & 90.6 & 88.9 & 74.5 & 82.9 & \multirow{2}{*}{-} \\
& DiagnosisGAN & 73.7 & 86.7 & \textbf{92.0} & 89.4 & \textbf{75.0} & \textbf{83.4} &  \\
\hline\hline
Four-phase & Cls-4P & 73.3 & 87.0 & 91.3 & 91.4 & 74.2 & 83.5 & - \\
\hline
\end{tabular}
\end{center}
\label{tab:main_comp}
\end{table*}
\subsection{Baselines and evaluation metric}
\noindent \textbf{Baselines.}
We compare our method with the following baselines:
\begin{itemize}
    \item \textbf{Classification with three-phase CT (Cls-3P).} A naive approach for diagnosis with missing data is to build a model that learns to classify cancer subtypes using the given three-phase CT. As there are four possible combinations of phases for three-phase CT, we train a separate classifier for each combination independently, resulting in four distinct cancer subtype classifiers. We explore whether a joint synthesis and classification approach can performe better than this naive baseline for diagnosis with incomplete data.
    \item \textbf{CollaGAN.} We use the multiple cycle consistency loss for the reconstruction of multiple input images as proposed in CollaGAN~\cite{Lee2019collagan}, which performed better on missing MRI contrast synthesis than existing single input image-based synthesis methods~\cite{choi2018stargan, zhu2017cyclegan}. We predict cancer subtypes using CT phases generated by this baseline and compare the diagnostic performance with our results.
    \item \textbf{Synthesis with segmentation (Syn-Seg).} The use of the segmentation network $S$ as extra supervision on the generator $G$ is proposed in~\cite{huo2019synseg, zhang2018trans}, which aims at producing synthetic CT images that are effective for the segmentation task. We investigate the effect of this segmentation supervision on the performance of diagnosis with synthesized CT images, which has not yet been well explored in the previous works.
    \item \textbf{ReMIC.} We extract the shared content features across phases and reconstruct images using the extracted content feature and phase-specific style representation, following the method of~\cite{tmi2021comp}. To incorporate the style representation in the image generation, the adaptive instance normalization (AdaIN)~\cite{AdaIN} is used for the normalization layer.
\end{itemize}

Since baseline synthesis methods only generate a 2D slice image but not a 3D CT volume, we implement baselines based on 3D volume generator architecture used in our method. We apply the key components of each baseline to the generator to compare their effect on the downstream diagnosis task. We have trained and tuned all the baselines in a similar way with the proposed method to obtain the optimal performance for each method.

\noindent \textbf{Evaluation metric.}
To evaluate whether the synthesized images are diagnostically useful, we use the area under the receiver operating characteristic curve (AUC) to measure the cancer subtype classification performance. The AUC score for each subtype class is measured using a one-vs-all strategy, and the mean AUC (mAUC) is computed by averaging AUC among all classes. The cancer subtype prediction result is computed from the completed four-phase CT using the classifier $C$ trained with the original complete CT dataset. For the Cls-3P baseline, the subtype prediction result for the given three-phase CT is obtained by the trained classifier for the corresponding phase combination. The higher AUC value indicates that the generated CT images are more effective for the subtype classification. To evaluate the statistical significance of the performance gain of our method, we perform a permutation test with 10,000 permutations and compute \textit{p}-value. 
We compute voxel-to-voxel comparison metrics including peak-signal-to-noise ratio (PSNR) and structural similarity index measure (SSIM), but these voxel-wise image quality metrics are not expected to reflect the effectiveness of synthesized images for our target task of the cancer subtype classification. In addition, with the intuition that a more accurate tumor segmentation map can be estimated from a better synthesized image, we apply the segmentation model on the synthesized image and measure the Dice score.

\subsection{Ablation study}
We conduct an ablation study on the impact of hyper-parameters, $\lambda_{rec}$, $\lambda_{seg}$, and $\lambda_{cls}$ in (\ref{eqa:L_full}), on the learning process of the proposed DiagnosisGAN. We set $\lambda_{rec}=1$ by default, and then investigate the influence of the other parameters by setting $\lambda_{seg}\in\{0.1, 1.0\}$ and $\lambda_{cls}\in\{0.1, 1.0\}$. Fig.~\ref{fig:loss_curves_abl} depicts the loss curves of DiagnosisGAN for training and validation on different hyper-parameter settings.
When $\lambda_{seg}=1.0$ and $\lambda_{cls}=1.0$, $\mathcal{L}_{adv}^G$ does not well converge, while the desired behavior is $\mathcal{L}_{adv}^G=\mathcal{L}_{adv}^D=0.25$ as we adopted the least-square adversarial loss. 
For $\lambda_{seg}=0.1$ and $\lambda_{cls}=1.0$, $\mathcal{L}_{seg}$ and $\mathcal{L}_{adv}^G$ do not converge.
Therefore, we consider the hyperparameter setting shows the better convergence if $\mathcal{L}_{adv}^G=\mathcal{L}_{adv}^D=0.25$ and $\mathcal{L}_{seg}$ and $\mathcal{L}_{cls}$ gradually decrease with less fluctuation through the training epochs.
The experimental results show that the convergence behavior for $\lambda_{seg}=0.1$, $\lambda_{cls}=0.1$ is more stable than the other settings.




\subsection{Quantitative results}
We compare the performance of our DiagnosisGAN with the baselines for incomplete CT diagnosis. For each complete four-phase CT scan in the test dataset, four simulated incomplete CT scans are generated as described in Section~\ref{sec:dataset}, resulting in 37$\times$4 simulated test cases. We measure the cancer subtype classification performance for all produced completed CT scans on this set. 

The results are summarized in Table~\ref{tab:main_comp}. ``Cls-4P'' refers to the subtype classification on original four-phase CT using the classifier $C$ trained with complete data. It is desired to achieve the performance close to this setting by generating informative samples for incomplete CT scans. It can be seen that the mAUC of Cls-3P is lower than Cls-4P, which is an expected result as Cls-3P lacks some visual information for characterizing tumors. For ccRCC, Cls-3p was better than Cls-4p, which means that ccRCC can be better classified using only three phases. ``BaseSyn'' means the generator $G$ is trained with only the adversarial loss $\mathcal{L}_{adv}^G$ and the reconstruction loss $\mathcal{L}_{rec}$, which serves as a starting point for missing phase synthesis. A lower mAUC is observed for BaseSyn compared to Cls-3P, indicating that the synthesized images do not benefit the subsequent classification task. Adding the multiple cycle consistency loss of CollaGAN to the training loss of the generator $G$ does not bring a performance gain over BaseSyn in terms of AUC. The representational disentanglement of ReMIC also does not show an improvement in classification performance. Supervising the generator $G$ using the segmentation network $S$ (Syn-Seg) improves the classification performance to some degree (0.2\% mAUC). However, there is still a large gap between these baselines and Cls-4P.


\begin{table}
\caption{Quantitative comparison of cancer subtype classification performances with the full test set (37$\times$4 simulated + 66 real incomplete cases).}
\begin{center}
\begin{tabular}{|l|l|c|c|}
\hline 
Classifier input & Method & mAUC & $p$-value\\
\hline\hline
Three-phase & Cls-3P & 78.1 & 0.0227 \\
\hline
\multirow{5}{7em}{Three-phase + One synthesized} & BaseSyn  & 77.8 & 0.0053 \\
 & CollaGAN & 77.3 & 0.0049\\
 &Syn-Seg & 77.9 & 0.0080  \\
 &ReMIC & 77.7 & 0.0029 \\
\cline{2-4}
& DiagnosisGAN & \textbf{79.6} & -\\
\hline
\end{tabular}
\end{center}
\label{tab:main_comp_real}
\end{table}
\begin{table}[]
\caption{Quantitative comparison on 139 incomplete multi-phase CT cases from TCIA.}
\begin{center}
\begin{tabular}{|l|l|c|c|}
\hline 
Classifier input & Method & mAUC & $p$-value\\
\hline\hline
Three-phase & Cls-3P & 83.6 & 0.0738\\
\hline
\multirow{6}{7em}{Three-phase + One synthesized} & BaseSyn & 83.4 & 0.0704 \\
& CollaGAN & 83.2 & 0.0379 \\
& Syn-Seg & 83.5 & 0.1656 \\
& ReMIC & 82.0 & 0.0288\\
\cline{2-4}
& DiagnosisGAN & \textbf{85.9} & -\\
\hline
\end{tabular}
\end{center}
\label{tab:main_comp_tcia}
\end{table}


On the other hand, the proposed DiagnosisGAN without joint learning (Diagnosis w/o joint) achieves the mean AUC of 82.9\%, surpassing all the baselines by a large margin. Here the generator $G$ is trained using the full objective $\mathcal{L}$ with the fixed classifier $C$. The performance of DiagnosisGAN is further boosted by joint optimization of the generator $G$ and the classifier $C$, resulting in the performance gain of 0.5\%. We confirm that the mAUC gains of our DiagnosisGAN over all the baseline methods are statistically significant (\textit{p}-value$<$0.05). Our approach achieves better AUC than Cls-4p in some of the classes. Since we jointly optimize the generator and the classifier, the generator learns to synthesize images that can lead to accurate classification.


We further evaluate our DiagnosisGAN with the baselines on the full test set containing both simulated and 66 ``real'' incomplete three-phase CT cases. All models including baselines and DiagnosisGAN have the same parameters used in the evaluation in Table~\ref{tab:main_comp}. Note that the performance of Cls-4P cannot be measured on this set since the ground-truth CT image for the missing phase is not available in a real incomplete CT case. The  results are summarized Table~\ref{tab:main_comp_real}. We can observe a similar trend that DiagnosisGAN outperforms all the baselines with statistical significance (\textit{p}-value$<$0.05). 



To explore the generalizability of our approach, we conduct experiments on an independent test dataset from TCIA. Since the TCIA dataset only contains three RCC subtype classes, we first train a three-class classifier using RCC data in our in-house dataset, and then evaluate the classification performance of the models on the TCIA dataset. Note that in this experiment, DiagnosisGAN is trained with the three-class classifier. The results are summarized in Table~\ref{tab:main_comp_tcia}. We can see that DiagnosisGAN still achieves the highest performance among all compared methods, which shows that our joint learning method can generalize to external data.

These results demonstrate that our method produces more diagnostically useful CT images than existing synthesis approaches so that it is much better to classify cancer subtypes with the completed four-phase CT rather than the given three-phase CT.


\begin{figure*}[!t]
\begin{center}
\includegraphics[width=1.0\linewidth]{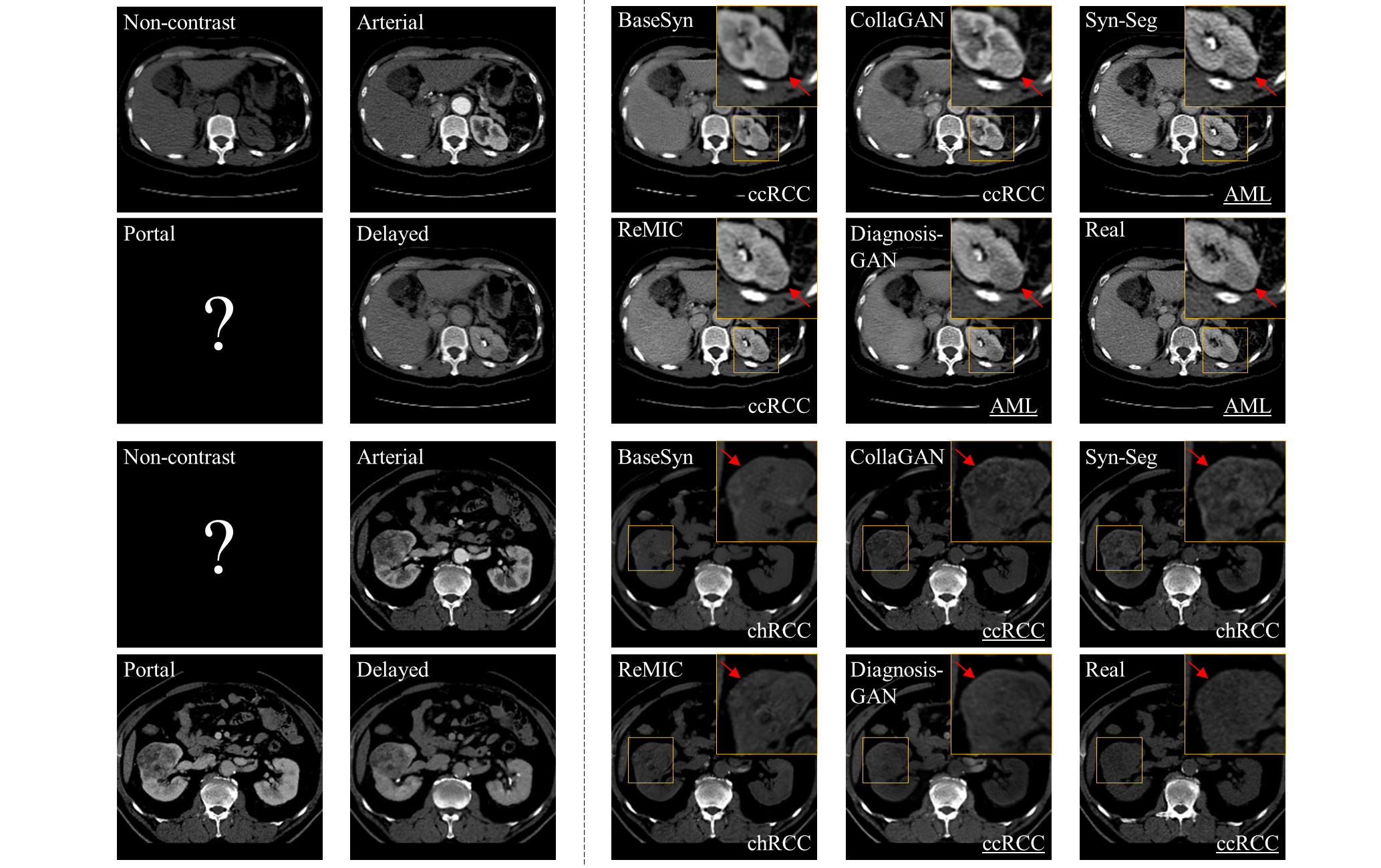}
\end{center}
   \caption{Qualitative comparison. Tumor regions are zoomed-in and indicated by red arrows. The actual cancer subtype labels are shown in the real image. The predicted subtype class using the completed four-phase CT is displayed on each result, where the correct labels are underlined.
   }
\label{fig:Comparison}
\end{figure*}

\begin{figure*}[!t]
\begin{center}
\includegraphics[width=1.0\linewidth]{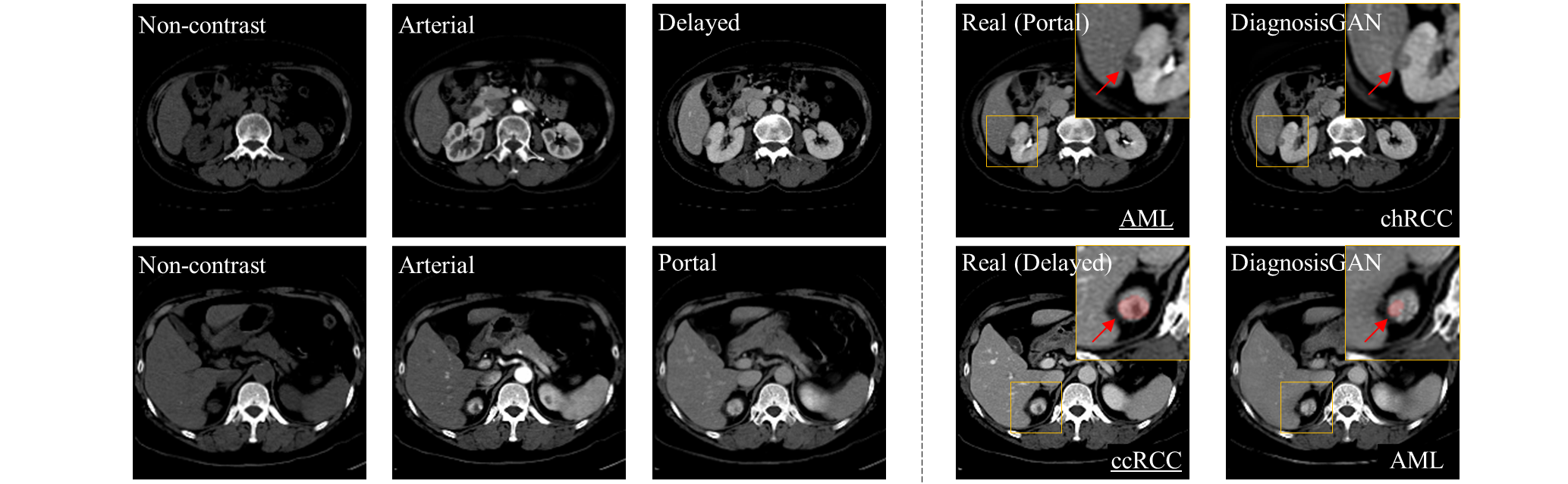}
\end{center}
   \caption{Qualitative examples of failure cases. For bottom bottom example, the ground-truth tumor segmentation mask and the predicted segmentation mask are overlaid with zoomed-image.
   }
\label{fig:fig_failure}
\end{figure*}

\begin{table}
\caption{Evaluation of synthesized image quality in terms of PSNR, SSIM, and Dice score (mean$\pm$std).}
\begin{center}
{\small
\begin{tabular}{|l|c|c|c|}
\hline 
Method & PSNR & SSIM & Dice (\%) \\
\hline
BaseSyn & 20.14$\pm$1.98 & 0.6188$\pm$0.1009 & 74.0$\pm$24.1 \\
CollaGAN & 20.01$\pm$2.15 & 0.6221$\pm$0.1065 & 74.6$\pm$24.5 \\
Syn-Seg & 19.69$\pm$2.11 & 0.6123$\pm$0.1051 & 78.0$\pm$21.0 \\
ReMIC & 20.33$\pm$2.02 & 0.6333$\pm$0.1012 & 75.2$\pm$24.4 \\
DiagnosisGAN & 20.07$\pm$2.06 & 0.6243$\pm$0.1045 & 77.9$\pm$20.8 \\
\hline
\end{tabular}
}
\end{center}
\label{tab:image_metric}
\end{table}

\begin{table*}[!t]
\caption{Quantitative performance comparison of missing CT phase synthesis and cancer subtype classification for each missing phase case. The bolded numbers represent the best results, and the underline indicates the second best.}
\begin{center}
{\small
\begin{tabular}{|l|c c c|c c c|c c c|c c c|}
\hline 
\multirow{2}{7em}{Method} & \multicolumn{3}{|c|}{Non-contrast}  & \multicolumn{3}{|c|}{Arterial}  & \multicolumn{3}{|c|}{Portal} & \multicolumn{3}{|c|}{Delayed}  \\
 & SSIM & Dice & mAUC & SSIM & Dice & mAUC & SSIM & Dice & mAUC & SSIM & Dice & mAUC\\
\hline
BaseSyn & \underline{0.6155} & 57.9  & 80.6  & 0.6284 & 80.1 & \underline{80.0} & 0.6287 & 75.6 & 80.0 & 0.6027  & \underline{82.5} & 81.1 \\
CollaGAN & 0.5974 & 61.5 & \textbf{82.0}  & 0.6364 & 78.7 & 79.6 & \textbf{0.6442}  & 76.4 & 80.1 & 0.6102  & 81.6 & 80.9 \\
Syn-Seg & 0.6097 & \textbf{66.5} & 77.6  & 0.6414 & \underline{80.2} & 78.3 & 0.5931 & \underline{82.8} & 80.2 & 0.6019  & \textbf{82.6} & \underline{81.3} \\
ReMIC & \textbf{0.6260} & 61.0 & 79.1  & \underline{0.6453} & 78.7  & 79.2 & 0.6252 & 81.3 & \underline{80.3} & \textbf{0.6368}  & 79.8 & 81.0 \\
DiagnosisGAN  & 0.6105 & \underline{64.0} & \underline{81.7}  & \textbf{0.6549} & \textbf{82.2}  & \textbf{82.0} & \underline{0.6408} & \textbf{83.0} & \textbf{83.1} & \underline{0.6210}  & 82.4 & \textbf{83.4} \\
\hline
\end{tabular}
}
\end{center}
\label{tab:phase_eval}
\end{table*}

\subsection{Evaluation of synthesized image quality}
Fig.~\ref{fig:Comparison} shows several examples where the proposed DiagnosisGAN leads to correct cancer subtype classification. Here we present example results from BaseSyn, CollaGAN, Syn-Seg, ReMIC, and DiagnosisGAN for different missing phase cases. For each case, the true cancer subtype label is shown in the real image. The tumor areas are zoomed in for more details. We predict the cancer subtype probabilities $\hat{p}$ from the completed four-phase CT $\mathcal{I}_m^+$ using the classifier $C$, and the subtype class having the highest probability in $\hat{p}$ is displayed in each synthesized result ${\hat{I}}_m$. As can be seen, with the synthesized images of DiagnosisGAN, the cancer subtypes are correctly classified in all cases, while CollaGAN and Syn-Seg lead to misclassification in several cases. For example, in the case of synthesizing portal phase (the first two rows), BaseSyn, CollaGAN and ReMIC misclassify the cancer subtypes as ccRCC, where the true label is AML, while DiagnosisGAN yields the accurate subtype prediction result. We can see that the tumor regions synthesized by DiagnosisGAN have a similar visual appearance to that of the real images, while the results of the baseline methods show different image characteristics around tumor regions, such as intensity and contrast, and contain some artifacts that are not present in the real images. It has to be noted that it is nontrivial to judge the diagnostic usefulness of the synthesized CT images solely by visual inspection since renal tumors exhibit subtle differences in image features across cancer subtypes, leading to inter-observer variability~\cite{deepkidney, subtle}. However, we can clearly see that synthesizing missing phases by using DiagnosisGAN is beneficial for the cancer subtype classification of hard samples, which are misclassified by the baselines. 
We also visualize some failed cases in Fig.~\ref{fig:fig_failure}.
For the top example in Fig. ~\ref{fig:fig_failure}, the size of the renal mass is 1cm, which is too small to accurately differentiate tumor subtypes in CT. The true subtype of tumor is AML, but it is misclassified as chRCC by our method. 
Besides, for the bottom example, the inaccurate segmentation of tumor boundaries could cause misclassification, but this issue can be alleviated by performing further refinement process.

Table~\ref{tab:image_metric} presents the evaluation of synthesized image quality in terms of PSNR, SSIM, and Dice score. It can be seen that our DiagnosisGAN achieves high performance in terms of both voxel-wise metrics (PSRN/SSIM) and Dice score. Syn-Seg and DiagnosisGAN achieve much higher Dice scores than other baselines, which means that the supervision of tumor segmentation can lead to well-defined tumor structure in synthesized images. It should be noted that better performance on voxel-wise image quality metrics does not guarantee better diagnostic performance. Importantly, our proposed method leads to large improvement of the end task performance, i.e., cancer subtype classification, over the baselines, as can be seen in Table~\ref{tab:main_comp}, ~\ref{tab:main_comp_real}, and~\ref{tab:main_comp_tcia}.
\\

\subsection{Phase-level analysis}
To investigate the effectiveness of our model on the phase level, we further break down the performance comparison between our method and baselines into individual missing phase case.
We report SSIM, Dice score, and mAUC of all methods for each missing phase case.
The results are summarized in Table~\ref{tab:phase_eval}. We can find that when arterial phase is missing, our DiagnosisGAN achieves the best performance on all metrics, which demonstrate the effectiveness of our framework for both missing phase synthesis and cancer subtype classification.
When portal phase is missing, DiagnosisGAN achieves the best Dice and mAUC, and the second best SSIM scores.
Specifically, in terms of mAUC, our model surpasses second best performing method by a significant margin of 2.8\%.
The highest cancer subtype classification performance is obtained when delayed phase is missing.


\begin{table}[!t]
\caption{Quantitative results of cancer subtype classification with manual segmentation maps.}
\begin{center}
\begin{tabular}{|l|l|c|c|}
\hline 
Classifier input & Method & mAUC & $p$-value\\
\hline\hline
Three-phase & Cls-3P & 87.8 & 0.0107\\
\hline
\multirow{6}{7em}{Three-phase + One synthesized} & BaseSyn & 87.7 & 0.0060 \\
& CollaGAN  & 87.7 & 0.0044 \\
& Syn-Seg  & 87.3 & 0.0017\\
& ReMIC & 87.6 & 0.0144\\
\cline{2-4}
& DiagnosisGAN (w/o joint)   & 89.7 & \multirow{2}{*}{-}\\
& DiagnosisGAN   & \textbf{90.0} & \\
\hline\hline
Four-phase & Cls-4P  & 91.8 & -\\
\hline
\end{tabular}
\end{center}
\label{tab:seg_comp}
\end{table}

\subsection{Impact of segmentation quality}
Since the segmentation map $\hat{s}$ predicted by the segmentation network $S$ is used to produce the lesion-level features $f$ through masked average pooling for classifying cancer subtypes, the performance of the classifier $C$ depends on the quality of the output segmentation map $\hat{s}$. It is expected that the classification performance will be improved with more accurate segmentation maps. To investigate how much the quality of predicted segmentation influences the overall classification performance, we use the ground-truth (manually annotated) segmentation map $s$ instead of the prediction output $\hat{s}$ for lesion-level feature extraction, that is, $\hat{s}$ is replaced by $s$ in (\ref{eqa:MAP}). Under this setting, the classifier $C$ for complete four-phase CT is trained and tested, and then used to evaluate all synthesis methods. As the classifier $C$ involves the training of DiagnosisGAN, we train DiagnosisGAN again in this setting, while the baseline models remain the same as the previous sub-section.

The results are summarized in Table~\ref{tab:seg_comp}. As expected, the mAUCs for all methods are much higher than the results obtained using the predicted segmentations, which are presented in Table~\ref{tab:main_comp}. For example, there is a gap of 8.3\% for Cls-4P (83.5\% vs 91.8\%) and 6.3\% for Cls-3P (81.5\% vs 87.8\%). These differences indicate the impact of the segmentation quality on the performance of cancer subtype classification in our experiments. Enhancing the performance of the segmentation network $S$ may benefit the classification task, which is beyond the scope of this paper. Note that on the test cases, the average Dice scores of the segmentation network for kidney and tumor are 0.969$\pm$0.014 and 0.856$\pm$0.131, respectively.  

When comparing our method to the baselines, we observe similar trends, that is, DiagnosisGAN w/o joint significantly improves the performance of BaseSyn by 2.0\% mAUC, and DiagnosisGAN outperforms all the baselines by a large margin with statistical significance (\textit{p}-value$<$0.05). The joint learning strategy further boosts the performance by 0.3\% mAUC.
We can see that Syn-Seg does not bring a performance gain over BaseSyn, unlike the result of the previous experiment. This can be interpreted that the supervision of the segmentation task helps improve the quality of the predicted segmentation map from the synthesized image, but it does not provide useful information for the differentiation of renal tumors. These results demonstrate the necessity and effectiveness of supervising the learning of the missing phase synthesis with cancer subtype classification.


\begin{table}
\caption{Quantitative comparison of cancer subtype classification performances when two CT phases are missing ($p$-value $<$ 0.05).}
\begin{center}
\begin{tabular}{|l|l|c|}
\hline 
Classifier input & Method & mAUC\\
\hline\hline
Two-phase & Cls-2P & 79.8 \\
\hline
\multirow{5}{7em}{Two-phase + Two synthesized} & BaseSyn  & 77.2  \\
 & Syn-Seg & 77.6 \\
 &ReMIC  & 75.4  \\
\cline{2-3}
& DiagnosisGAN (w/o joint) & 81.0  \\
& DiagnosisGAN & \textbf{81.6} \\
\hline\hline
Three-phase & Cls-3P & 81.5\\
\hline
\end{tabular}
\end{center}
\label{tab:two_missing}
\end{table}

\subsection{Cases of more than one missing phase}
As mentioned in Section~\ref{sec:Method}, our method can be applied to the cases with more than one missing phase. In these cases, if there are $N_m$ missing phases, the output of the generator has $N_m$ channel dimension, where each channel of the output is a synthesized image for the corresponding missing phase index. We conduct experiments to investigate the performance of DiagnosisGAN for the case of multiple missing phases. Table~\ref{tab:two_missing} shows the results for the case of two missing phases. It can be seen that the mAUC of classification with two-phase CT (Cls-2P) is lower than Cls-3P as expected. We can observe that DiagnosisGAN outperforms all the baselines by a large margin with statistical significance (\textit{p}-value$<$0.05). Interestingly, DiagnosisGAN achieves similar performance to Cls-3P, which indicates the effectiveness of our method for the diagnosis with missing phases.

Table~\ref{tab:three_missing} shows the results for the case of three missing phases. As can be seen, classification with given one-phase CT (Cls-1P) shows much lower performance than Cls-2P, which is as expected since generating three missing phases from one CT phase image is a very challenging task. Meanwhile, our DiagnosisGAN still outperforms all the baselines by a large margin with statistical significance (\textit{p}-value$<$0.05).

\begin{table}
\caption{Quantitative comparison of cancer subtype classification performances when three CT phases are missing ($p$-value $<$ 0.05).}
\begin{center}
\begin{tabular}{|l|l|c|}
\hline 
Classifier input & Method & mAUC\\
\hline\hline
One-phase & Cls-1P & 73.8 \\
\hline
\multirow{5}{7.5em}{One-phase + Three synthesized} & BaseSyn  & 68.0  \\
 & Syn-Seg & 68.2 \\
 &ReMIC  & 69.2  \\
\cline{2-3}
& DiagnosisGAN (w/o joint) & 75.0  \\
& DiagnosisGAN & \textbf{75.4} \\
\hline\hline
Two-phase & Cls-2P & 79.8\\
\hline
\end{tabular}
\end{center}
\label{tab:three_missing}
\end{table}

\section{Conclusion}
\label{conclusion}
In this paper, we propose a novel joint learning framework for missing CT phase synthesis and cancer subtype classification, termed DiagnosisGAN, which learns to generate missing images that are effective for classifying pathological subtypes of the tumor. Extensive experiments on 249 in-house and 139 external multi-phase CT scans of kidney cancer patients demonstrate the effectiveness and superiority of our framework over the baselines for the diagnosis with missing data. Notably, with the images generated by DiagnosisGAN, we even achieve comparable performance to the diagnosis with real CT images. We believe that our work can serve as a strong baseline for future research in diagnostically meaningful medical image synthesis. Our future work will focus on the case of more than one missing phase and involve diagnosing lesions of other organs, such as the liver.


\begin{thebibliography}{00}

\bibitem{kim2002rcc} J. K. Kim, T. K. Kim, H. J. Ahn, C. S. Kim, K.-R. Kim, and K.-S. Cho, ''Differentiation of subtypes of renal cell carcinoma on helical CT scans," \emph{Am. J. Roentgenol.}, vol. 178, pp. 1499–1506, 2002.

\bibitem{wang2021rcc} X. Wang, G. Song, and H. Jiang, ``Differentiation of renal angiomyolipoma without visible fat from small clear cell renal cell carcinoma by using specific region of interest on contrast-enhanced ct: a new combination of quantitative tools,” \emph{Cancer Imaging}, vol. 21, no. 47, 2021

\bibitem{han2019rcc} S. Han, S. I. Hwang, and H. J. Lee, ``The classification of renal cancer in 3-phase CT images using a deep learning method,” \emph{J. Digit. Imaging}, vol. 32, no. 4, pp. 3660–3672, 2019.

\bibitem{young2013ccrcc} J. R. Young, D. Margolis, S. Sauk, A. J. Pantuck, J. Sayre, and S. S. Raman, ``Clear cell renal cell carcinoma: Discrimination from other renal cell carcinoma subtypes and oncocytoma at multiphasic multidetector ct,”\emph{Radiology}, vol. 267, no. 2, pp. 444–453, 2013.

\bibitem{drevelegas2010brain} A. Drevelegas and N. Papanikolaou, \emph{Imaging Modalities in Brain
Tumors.} Springer Berlin Heidelberg, 2011, pp. 13–33.

\bibitem{CHASE} A. Raju, C.-T. Cheng, Y. Huo, J. Cai, J. Huang, J. Xiao, L. Lu, C. Liao, and A. P. Harrison, ``Co-heterogeneous and adaptive segmentation from multi-source and multi-phase ct imaging data: A study on pathological liver and lesion segmentation,” in \emph{Proceedings of the European Conference on Computer Vision}, 2020, pp. 448–465.

\bibitem{breath} J. R. Mains, F. Donskov, E. M. Pedersen, H. H. T. Madsen, and F. Rasmussen, ``Dynamic contrast-enhanced computed tomography as a potential biomarker in patients with metastatic renal cell carcinoma preliminary results from the danish renal cancer group study-1,” \emph{Investig. Radiol.}, vol. 49, no. 9, pp. 601–607, 2014.

\bibitem{incidental} A. Volpe, T. Panzarella, R. A. Rendon, M. A. Haider, F. I. Kondylis, and M. A. S. Jewett, ``The natural history of incidentally detected small renal masses,” \emph{Cancer}, vol. 100, no. 4, pp. 738–745, 2004.

\bibitem{radiation} D. J. Brenner and E. J. Hall, ``Computed tomography —an increasing source of radiation exposure,” \emph{N. Engl. J. Med}, vol. 357, no. 22, pp.
2277–2284, 2007.

\bibitem{cost} J. Chen, J.Wei, and R. Li, ``TarGAN: Target-aware generative adversarial networks for multi-modality medical image translation,” in \emph{Proceedings of the Medical Image Computing and Computer-Assisted Intervention}, 2021.

\bibitem{su2015multi} H. Su, S. Maji, E. Kalogerakis, and E. Learned-Miller, ``Multi-view convolutional neural networks for 3d shape recognition,” in \emph{Proceedings of the IEEE international conference on computer vision}, 2015, pp. 945–953.

\bibitem{YUAN2012} L. Yuan, Y. Wang, P. M. Thompson, V. A. Narayan, and J. Ye, ``Multi-source feature learning for joint analysis of incomplete multiple heterogeneous neuroimaging data,” \emph{NeuroImage}, vol. 61, no. 3, pp. 622– 632, 2012.

\bibitem{Zhou2019} T. Zhou, M. Liu, K.-H. Thung, and D. Shen, ``Latent representation learning for alzheimer’s disease diagnosis with incomplete multimodality neuroimaging and genetic data,” \emph{IEEE Transactions on Medical Imaging}, vol. 38, no. 10, pp. 2411–2422, 2019.

\bibitem{Lee2019collagan} D. Lee, J. Kim, W.-J. Moon, and J. C. Ye, ``CollaGAN: Collaborative GAN for missing image data imputation,” in \emph{Proceedings of the IEEE Conference on Computer Vision and Pattern Recognition}, June 2019, pp. 2487–2496.

\bibitem{choi2018stargan} Y. Choi, M. Choi, M. Kim, J.-W. Ha, S. Kim, and J. Choo, ``StarGAN: Unified generative adversarial networks for multi-domain image-toimage translation,” in \emph{Proceedings of the IEEE Conference on Computer Vision and Pattern Recognition}, June 2018, pp. 8789–8797.

\bibitem{goodfellow2014gan} I. Goodfellow, J. Pouget-Abadie, M. Mirza, B. Xu, D. Warde-Farley, S. Ozair, A. Courville, and Y. Bengio, ``Generative adversarial nets,” in \emph{Proceedings of Advances in Neural Information Processing Systems}, vol. 27, 2014, pp. 2672–2680.

\bibitem{isola2017pix2pix} P. Isola, J.-Y. Zhu, T. Zhou, and A. A. Efros, ``Image-to-image translation with conditional adversarial networks,” in \emph{Proceedings of the IEEE Conference on Computer Vision and Pattern Recognition}, July 2017, pp. 1125–1134.

\bibitem{seo2021neural} M. Seo, D. Kim, K. Lee, S. Hong, J. S. Bae, J. H. Kim, and S. Kwak, ``Neural contrast enhancement of CT image,” in \emph{Proceedings of the IEEE Winter Conference on Applications of Computer Vision}, Jan.2021, pp. 3973–3982.

\bibitem{zhu2017cyclegan} J.-Y. Zhu, T. Park, P. Isola, and A. A. Efros, ``Unpaired image-to-image translation using cycle-consistent adversarial networks,” in \emph{Proceedings of the IEEE International Conference on Computer Vision}, 2017, pp. 2223–2232.

\bibitem{zhang2018trans} Z. Zhang, L. Yang, and Y. Zheng, ``Translating and segmenting multimodal medical volumes with cycle- and shape-consistency generative adversarial network,” in \emph{Proceedings of the IEEE Conference on Computer Vision and Pattern Recognition}, 2018, pp. 9242–9251.

\bibitem{choi2020starganv2} Y. Choi, Y. Uh, J. Yoo, and J.-W. Ha, ``StarGAN v2: Diverse image synthesis for multiple domains,” in \emph{Proceedings of the IEEE Conference on Computer Vision and Pattern Recognition}, June 2020, pp. 8188–8197.

\bibitem{huang2018munit} X. Huang, M.-Y. Liu, S. Belongie, and J. Kautz, ``Multimodal unsupervised image-to-image translation,” in \emph{Proceedings of the European Conference on Computer Vision}, 2018, pp. 179–196.

\bibitem{kim2017discogan} T. Kim, M. Cha, H. Kim, J. K. Lee, and J. Kim, ``Learning to discover cross-domain relations with generative adversarial networks,” in \emph{Proceedings of the International Conference on Machine Learning}, vol. 70, 2017, pp. 1857–1865.

\bibitem{zhu2017bicyclegan} J.-Y. Zhu, R. Zhang, D. Pathak, T. Darrell, A. A. Efros, O. Wang, and E. Shechtman, ``Toward multimodal image-to-image translation,” in \emph{Proceedings of the Advances in Neural Information Processing Systems}, vol. 30, 2017, pp. 465–476.

\bibitem{yoon2018radialgan} J. Yoon, J. Jordon, and M. van der Schaar, ``RadialGAN: Leveraging multiple datasets to improve target-specific predictive models using generative adversarial networks,” in \emph{Proceedings of the International Conference on Machine Learning}, vol. 80, 2018, pp. 5699–5707.

\bibitem{FRIDADAR2018321} M. Frid-Adar, I. Diamant, E. Klang, M. Amitai, J. Goldberger, and H. Greenspan, ``Gan-based synthetic medical image augmentation for increased cnn performance in liver lesion classification,” \emph{Neurocomputing}, vol. 321, pp. 321–331, 2018.

\bibitem{focal_aug} L. H, L. H, H. H, B. H, L. JS, and K. J., ``Classification of focal liver lesions in ct images using convolutional neural networks with lesion information augmented patches and synthetic data augmentation,” \emph{Med. Phys.}, vol. 321, pp. 5029–5046, 2021.

\bibitem{lung_aug} ——, ``Synthetic ct image generation of shape-controlled lung cancer using semi-conditional infogan and its applicability for type classification,” \emph{Int. J. Comput. Assist. Radiol. Surg.}, vol. 16, pp. 241–251, 2021.

\bibitem{dar2019mri} S. U. Dar, M. Yurt, L. Karacan, A. Erdem, E. Erdem, and T. C¸ ukur, ``Image synthesis in multi-contrast MRI with conditional generative adversarial networks,” \emph{IEEE Trans. Med. Imag.}, vol. 38, no. 10, pp. 2375–2388, 2019.

\bibitem{sharma2020mri} A. Sharma and G. Hamarneh, ``Missing MRI pulse sequence synthesis using multi-modal generative adversarial network,” \emph{IEEE Trans. Med. Imag.}, vol. 39, no. 4, pp. 1170–1183, 2020.

\bibitem{yu2019eagan} B. Yu, L. Zhou, L. Wang, Y. Shi, J. Fripp, and P. Bourgeat, ``Ea-GANs: Edge-aware generative adversarial networks for cross-modality MR image synthesis,” \emph{IEEE Trans. Med. Imag.}, vol. 38, no. 7, pp. 1750–1762, 2019.

\bibitem{tmi2021comp} L. Shen, W. Zhu, X. Wang, L. Xing, J. M. Pauly, B. Turkbey, S. A. Harmon, T. H. Sanford, S. Mehralivand, P. L. Choyke, B. J. Wood, and D. Xu, ``Multi-domain image completion for random missing input data,” \emph{IEEE Trans. Med. Imag.}, vol. 40, no. 4, pp. 1113–1122, 2021.

\bibitem{liu2020dye} J. Liu, Y. Tian, A. M. A˘gıldere, K. M. Haberal, M. Cos¸kun, C. Duzgol, and O. Akin, ``DyeFreeNet: Deep virtual contrast CT synthesis,” in \emph{Proceedings of the International Workshop on Simulation and Synthesis in Medical Imaging}, 2020, pp. 80–89.

\bibitem{huo2019synseg} Y. Huo, Z. Xu, H. Moon, S. Bao, A. Assad, T. K. Moyo, M. R. Savona, R. G. Abramson, and B. A. Landman, ``SynSeg-Net: Synthetic segmentation without target modality ground truth,” \emph{IEEE Trans. Med. Imag.}, vol. 38, no. 4, pp. 1016–1025, 2019.

\bibitem{nie2017mri_ct} D. Nie, R. Trullo, J. Lian, C. Petitjean, S. Ruan, Q. Wang, and D. Shen, ``Medical image synthesis with context-aware generative adversarial networks,” in \emph{Proceedings of the Medical Image Computing and Computer Assisted Intervention}, 2017, pp. 417–425.

\bibitem{Wolterink2017mri_ct} J. M. Wolterink, A. M. Dinkla, M. H. F. Savenije, P. R. Seevinck, C. A. T. van den Berg, and I. Iˇsgum, ``Deep MR to CT synthesis using unpaired data,” in \emph{Proceedings of the Simulation and Synthesis in Medical Imaging}, 2017, pp. 14–23.

\bibitem{Pierorazio2013multi_ct} P. M. Pierorazio, E. S. Hyams, S. Tsai, Z. Feng, B. J. Trock, J. K. Mullins et al., ``Multiphasic enhancement patterns of small renal masses ($\leq$ 4 cm) on preoperative computed tomography: Utility for distinguishing subtypes of renal cell carcinoma, angiomyolipoma, and oncocytoma,” \emph{Urology}, vol. 81, no. 6, pp. 1265–1272, 2013.

\bibitem{Schieda2020multi_ct} N. Schieda, K. Nguyen, R. E. Thornhill, M. D. F. McInnes, M. Wu, and N. James, ``Importance of phase enhancement for machine learning classification of solid renal masses using texture analysis features at multi-phasic CT,” \emph{Abdom. Radiol.}, vol. 45, p. 2786–2796, 2020.

\bibitem{You2019multi_ct} M.-W. You, N. Kim, and H. Choi, ``The value of quantitative CT texture analysis in differentiation of angiomyolipoma without visible fat from clear cell renal cell carcinoma on four-phase contrast-enhanced CT images,” \emph{Clin. Radiol.}, vol. 74, no. 7, pp. 547–554, 2019.

\bibitem{cao2020liver} S.-E. Cao, L.-Q. Zhang, S.-C. Kuang, W.-Q. Shi, B. Hu, S.-D. Xie et al., ``Multiphase convolutional dense network for the classification of focal liver lesions on dynamic contrast-enhanced computed tomography,” \emph{World J. Gastroenterol.}, vol. 26, pp. 3660–3672, 2020.

\bibitem{huo2020liver} Y. Huo, J. Cai, C.-T. Cheng, A. Raju, K. Yan, B. A. Landman, J. Xiao, L. Lu, C.-H. Liao, and A. P. Harrison, ``Harvesting, detecting, and characterizing liver lesions from large-scale multi-phase CT data via deep dynamic texture learning,” 2020.

\bibitem{liang2018liver} D. Liang, L. Lin, H. Hu, Q. Zhang, Q. Chen, Y. lwamoto, X. Han, and Y.-W. Chen, ``Combining convolutional and recurrent neural networks for classification of focal liver lesions in multi-phase CT images,” in \emph{Proceedings of the Medical Image Computing and Computer Assisted Intervention}, 2018.

\bibitem{zhou2021liver} J. Zhou, W. Wang, B. Lei, W. Ge, Y. Huang, L. Zhang, Y. Yan, D. Zhou, Y. Ding, J. Wu, and W. Wang, ``Automatic detection and classification of focal liver lesions based on deep convolutional neural networks: A preliminary study,” \emph{Front. Oncol.}, vol. 10, p. 3261, 2021.

\bibitem{coy2019renal} H. Coy, K. Hsieh, W. Wu, M. B. Nagarajan, J. R. Young, M. L. Douek, M. S. Brown, F. Scalzo, and S. S. Raman, ``Deep learning and radiomics: the utility of Google TensorFlow™ Inception in classifying clear cell renal cell carcinoma and oncocytoma on multiphasic CT,” \emph{Abdom. Radiol.}, vol. 44, pp. 2009–2020, 2019.

\bibitem{oberai2020renal} A. Oberai, B. Varghese, S. Cen, T. Angelini, D. Hwang, I. Gill, M. Aron, C. Lau, and V. Duddalwar, ``Deep learning based classification of solid lipid-poor contrast enhancing renal masses using contrast enhanced CT,” \emph{Br. J. Radiol. Suppl.}, vol. 93, no. 1111, p. 20200002, 2020.

\bibitem{tanaka2020renal} T. Tanaka, Y. Huang, Y. Marukawa, Y. Tsuboi, Y. Masaoka, K. Kojima et al., ``Differentiation of small ($\leq$ 4 cm) renal masses on multiphase contrast-enhanced CT by deep learning,” \emph{Am. J. Roentgenol.}, vol. 214, no. 3, pp. 605–612, 2020.

\bibitem{deepkidney} K.-H. Uhm, S.-W. Jung, M. H. Choi, H.-K. Shin, J.-I. Yoo, S. W. Oh, J. Y. Kim, H. G. Kim, Y. J. Lee, S. Y. Youn, S.-H. Hong, and S.-J. Ko, ``Deep learning for end-to-end kidney cancer diagnosis on multi-phase abdominal computed tomography,” \emph{npj Precis. Onc.}, vol. 5, no. 54, 2021.

\bibitem{cicek2016unet3d} O¨ . C¸ ic¸ek, A. Abdulkadir, S. S. Lienkamp, T. Brox, and O. Ronneberger, ``3D U-Net: Learning dense volumetric segmentation from sparse annotation,” in \emph{Proceedings of the Medical Image Computing and Computer- Assisted Intervention}, 2016, pp. 424–432.

\bibitem{odena2016check} A. Odena, V. Dumoulin, and C. Olah, ``Deconvolution and checkerboard artifacts,” \emph{Distill}, 2016.

\bibitem{PANet} K. Wang, J. H. Liew, Y. Zou, D. Zhou, and J. Feng, ``PANet: Few-shot image semantic segmentation with prototype alignment,” in \emph{Proceedings of the IEEE International Conference on Computer Vision}, October 2019, pp. 9197–9206.

\bibitem{SGOne} X. Zhang, Y. Wei, Y. Yang, and T. S. Huang, ``SG-One: Similarity guidance network for one-shot semantic segmentation,” \emph{IEEE Trans. Cybern.}, vol. 50, no. 9, pp. 3855–3865, 2020.

\bibitem{LSGAN} X. Mao, Q. Li, H. Xie, R. Y. K. Lau, Z. Wang, and S. P. Smolley, ``Least squares generative adversarial networks,” in \emph{Proceedings of the IEEE International Conference on Computer Vision}, 2017, pp. 2813– 2821.

\bibitem{Dice} F. Milletari, N. Navab, and S. Ahmadi, ``V-net: Fully convolutional neural networks for volumetric medical image segmentation,” in \emph{Proceedings of the International Conference on 3D Vision}, 2016, pp. 565– 571.

\bibitem{EAU} B. Ljungberg, L. Albiges, Y. Abu-Ghanem, K. Bensalah, S. Dabestani, S. Fernandez-Pello et al., ``European association of urology guidelines on renal cell carcinoma: The 2019 update,” \emph{Eur. Urol.}, vol. 75, no. 5, pp. 799–810, 2019.

\bibitem{TCIA} K. Clark, B. Vendt, K. Smith, J. Freymann, J. Kirby, P. Koppel, S. Moore, S. Phillips, D. Maffitt, M. Pringle, L. Tarbox, and F. Prior, ``The cancer imaging archive (TCIA): Maintaining and operating a public information repository,” \emph{J. Digit. Imag.}, vol. 26, no. 6, pp. 1045–1057, 2013.

\bibitem{DEEDS} M. P. Heinrich, M. Jenkinson, M. Brady, and J. A. Schnabel, ``MRF based deformable registration and ventilation estimation of lung CT,” \emph{IEEE Trans. Med. Imag.}, vol. 32, no. 7, pp. 1239–1248, 2019.

\bibitem{Pytorch} A. Paszke, S. Gross, F. Massa, A. Lerer, J. Bradbury, G. Chanan et al., ``PyTorch: An imperative style, high-performance deep learning library,” in \emph{Proceedings of the Advances in Neural Information Processing Systems}, 2019, pp. 8024–8035.

\bibitem{IN} D. Ulyanov, A. Vedaldi, and V. S. Lempitsky, ``Instance normalization: The missing ingredient for fast stylization,” \emph{CoRR}, vol. abs/1607.08022, 2016.

\bibitem{LeakyRelu} A. L. Maas, A. Y. Hannun, and A. Y. Ng, ``Rectifier nonlinearities improve neural network acoustic models,” in \emph{Proceedings of the ICML Workshop on Deep Learning for Audio, Speech and Language Processing}, 2013.

\bibitem{Adam} D. P. Kingma and J. Ba, ``Adam: A method for stochastic optimization,” in \emph{Proceedings of the International Conference on Learning Representations}, 2015.

\bibitem{AdaIN} X. Huang and S. Belongie, ``Arbitrary style transfer in real-time with adaptive instance normalization,” in \emph{Proceedings of the IEEE International Conference on Computer Vision}, 2017, pp. 1501–1510.

\bibitem{subtle} S. A. Lee-Felker, E. R. Felker, N. Tan, D. J. A. Margolis, J. R. Young, J. Sayre, and S. S. Raman, ``Qualitative and quantitative MDCT features for differentiating clear cell renal cell carcinoma from other solid renal cortical masses,” \emph{Am. J. Roentgenol.}, vol. 203, no. 5, pp. 517–524, 2014.

\end{thebibliography}

\end{document}